\documentclass[aip,rsi,preprint,groupedaddress,showmsc,superscriptaddress,floatfix]{revtex4}
\usepackage{amsmath,amsfonts,amssymb,caption,hyperref,color,epsfig,graphics,graphicx,latexsym,mathrsfs,revsymb,theorem,url,verbatim,epstopdf}
\usepackage[a4paper]{geometry}
\usepackage[center]{titlesec}
\titleformat{\section}[hang]{\Large\bfseries\filcenter}{}{1em}{}
\titleformat{\subsection}[hang]{\bfseries}{}{1em}{}
\usepackage{amsmath,amsfonts,amssymb,caption,hyperref,color,epsfig,graphics,graphicx,latexsym,mathrsfs,revsymb,theorem,url,verbatim,epstopdf,cleveref}
\usepackage{braket}
\hypersetup{colorlinks,linkcolor={blue},citecolor={blue},urlcolor={red}}

\newtheorem{definition}{Definition}

\newtheorem{lemma}[definition]{Lemma}

\newtheorem{theorem}[definition]{Theorem}

\def\squareforqed{\hbox{\rlap{$\sqcap$}$\sqcup$}}
\def\qed{\ifmmode\squareforqed\else{\unskip\nobreak\hfil
\penalty50\hskip1em\null\nobreak\hfil\squareforqed
\parfillskip=0pt\finalhyphendemerits=0\endgraf}\fi}
\def\endenv{\ifmmode\;\else{\unskip\nobreak\hfil
\penalty50\hskip1em\null\nobreak\hfil\;
\parfillskip=0pt\finalhyphendemerits=0\endgraf}\fi}
\newenvironment{proof}{\noindent \textbf{{Proof.~} }}{\qed}
\def\Dbar{\leavevmode\lower.6ex\hbox to 0pt
{\hskip-.23ex\accent"16\hss}D}
\def\bpf{\begin{proof}}
\def\epf{\end{proof}}

\newcommand{\proj}[1]{|{#1}\rangle \langle {#1}|}

\newcommand{\nc}{\newcommand}

\def\bea{\begin{eqnarray}}
\def\eea{\end{eqnarray}}
\def\beq{\begin{equation}}
\def\eeq{\end{equation}}
\def\bal{\begin{aligned}}
\def\eal{\end{aligned}}
\def\bma{\begin{bmatrix}}
\def\ema{\end{bmatrix}}

\nc{\bbA}{\mathbb{A}} \nc{\bbB}{\mathbb{B}} \nc{\bbC}{\mathbb{C}}
\nc{\bbD}{\mathbb{D}} \nc{\bbE}{\mathbb{E}} \nc{\bbF}{\mathbb{F}}
\nc{\bbG}{\mathbb{G}} \nc{\bbH}{\mathbb{H}} \nc{\bbI}{\mathbb{I}}
\nc{\bbJ}{\mathbb{J}} \nc{\bbK}{\mathbb{K}} \nc{\bbL}{\mathbb{L}}
\nc{\bbM}{\mathbb{M}} \nc{\bbN}{\mathbb{N}} \nc{\bbO}{\mathbb{O}}
\nc{\bbP}{\mathbb{P}} \nc{\bbQ}{\mathbb{Q}} \nc{\bbR}{\mathbb{R}}
\nc{\bbS}{\mathbb{S}} \nc{\bbT}{\mathbb{T}} \nc{\bbU}{\mathbb{U}}
\nc{\bbV}{\mathbb{V}} \nc{\bbW}{\mathbb{W}} \nc{\bbX}{\mathbb{X}}
\nc{\bbZ}{\mathbb{Z}}

\nc{\bA}{{\bf A}} \nc{\bB}{{\bf B}} \nc{\bC}{{\bf C}}
\nc{\bD}{{\bf D}} \nc{\bE}{{\bf E}} \nc{\bF}{{\bf F}}
\nc{\bG}{{\bf G}} \nc{\bH}{{\bf H}} \nc{\bI}{{\bf I}}
\nc{\bJ}{{\bf J}} \nc{\bK}{{\bf K}} \nc{\bL}{{\bf L}}
\nc{\bM}{{\bf M}} \nc{\bN}{{\bf N}} \nc{\bO}{{\bf O}}
\nc{\bP}{{\bf P}} \nc{\bQ}{{\bf Q}} \nc{\bR}{{\bf R}}
\nc{\bS}{{\bf S}} \nc{\bT}{{\bf T}} \nc{\bU}{{\bf U}}
\nc{\bV}{{\bf V}} \nc{\bW}{{\bf W}} \nc{\bX}{{\bf X}}
\nc{\bZ}{{\bf Z}}

\nc{\bmA}{{\bm A}} \nc{\bmB}{{\bm B}} \nc{\bmC}{{\bm C}}
\nc{\bmD}{{\bm D}} \nc{\bmE}{{\bm E}} \nc{\bmF}{{\bm F}}
\nc{\bmG}{{\bm G}} \nc{\bmH}{{\bm H}} \nc{\bmI}{{\bm I}}
\nc{\bmJ}{{\bm J}} \nc{\bmK}{{\bm K}} \nc{\bmL}{{\bm L}}
\nc{\bmM}{{\bm M}} \nc{\bmN}{{\bm N}} \nc{\bmO}{{\bm O}}
\nc{\bmP}{{\bm P}} \nc{\bmQ}{{\bm Q}} \nc{\bmR}{{\bm R}}
\nc{\bmS}{{\bm S}} \nc{\bmT}{{\bm T}} \nc{\bmU}{{\bm U}}
\nc{\bmV}{{\bm V}} \nc{\bmW}{{\bm W}} \nc{\bmX}{{\bm X}}
\nc{\bmZ}{{\bm Z}}

\nc{\cA}{{\cal A}} \nc{\cB}{{\cal B}} \nc{\cC}{{\cal C}}
\nc{\cD}{{\cal D}} \nc{\cE}{{\cal E}} \nc{\cF}{{\cal F}}
\nc{\cG}{{\cal G}} \nc{\cH}{{\cal H}} \nc{\cI}{{\cal I}}
\nc{\cJ}{{\cal J}} \nc{\cK}{{\cal K}} \nc{\cL}{{\cal L}}
\nc{\cM}{{\cal M}} \nc{\cN}{{\cal N}} \nc{\cO}{{\cal O}}
\nc{\cP}{{\cal P}} \nc{\cQ}{{\cal Q}} \nc{\cR}{{\cal R}}
\nc{\cS}{{\cal S}} \nc{\cT}{{\cal T}} \nc{\cU}{{\cal U}}
\nc{\cV}{{\cal V}} \nc{\cW}{{\cal W}} \nc{\cX}{{\cal X}}
\nc{\cZ}{{\cal Z}}

\begin{document}
\title{Eigenvalues and eigenvectors of complex Hadamard matrices}

\author{Mengfan Liang}\email[]{lmf2021@buaa.edu.cn}
\affiliation{LMIB(Beihang University), Ministry of Education, and School of Mathematical Sciences, Beihang University, Beijing 100191, China}

\author{Lin Chen}\email[]{linchen@buaa.edu.cn(corresponding author)}
\affiliation{LMIB(Beihang University), Ministry of Education, and School of Mathematical Sciences, Beihang University, Beijing 100191, China}

\begin{abstract}
Characterizing the $6\times 6$ complex Hadamard matrices (CHMs) is an open problem in linear algebra and quantum information. In this paper, we investigate the eigenvalues and eigenvectors of CHMs. We show that any $n\times n$ CHM with dephased form has two constant eigenvalues $\pm\sqrt{n}$ and has two constant eigenvectors.  We obtain the maximum numbers of identical  eigenvalues of $6\times 6$ CHMs with dephased form and we extend this result to arbitrary dimension. We also show that there is no $6\times 6$ CHM with four identical eigenvalues. We conjecture that the eigenvalues and eigenvectors of $6\times 6$ CHMs will lead to the complete classification of $6\times 6$ CHMs. 
\end{abstract}

\date{\today}

\maketitle





\section{Introduction}
\label{sec:intro}
Mutually unbiased bases (MUB) is an important concept in quantum physics. 
Generally, 
 $k$ MUBs in $d$-dimensional Hilbert space $\bbC^d$ are $k$ orthogonal bases, such that any two vectors from different bases have inner product of modulus $1/\sqrt d$. Only when $k=d+1$, we say the  $k$ MUBs in $\bbC^d$ is complete. As far as we know, the complete MUBs exist in $\mathbb{C}^d$ when $d$ is a prime power \cite{WOOTTERS1989363}. How to find the complete MUBs in $\mathbb{C}^6$ is the first unsolved case and it is also a famous open problem in quantum information. The MUB problem has been studied in many ways. Paper \cite{rle11} investigated the average distance between four bases in six dimensions, and given a strong evidence that no four mutually unbiased bases exist in six dimensions. Paper \cite{jmm09} exhibited an infinite family of triplets of MUBs in dimension 6, and showed that this family cannot be extended to a complete MUBs in $\mathbb{C}^6$. Paper \cite{mw12jpa102001} showed that if a complete MUBs in $\mathbb{C}^6$ exists, then it cannot contain more than one product basis. Paper \cite{Chen2018Mutually} investigated the number of product vectors in the set of four MUBs in $\mathbb{C}^6$, and showed that in most cases the number of product vectors in each of the remaining three MUBs is at most two. More reaserch has been presented in \cite{Designolle2018Quantifying,mw12jpa135307,mpw16,Boykin05,bw08,bw10,deb10,wpz11,mw12ijqi,Chen2017Product}.\par
 The complex Hadamard matrix is also an important concept since it is usually related to MUB problem. An $n\times n$ matrix $H$ whose elements all have modulus one is called a CHM if $HH^{\dagger}=nI$. If a set of four MUBs in $\mathbb{C}^6$ exists and contains the identity, then any other matrix $U$ in the set satisfies that $\sqrt{6}U$ is a $6\times 6$ CHM, and we refer to the set as an MUB trio. The final target for us is to find if there exist an MUB trio. On the other hand, 
the complete classification of $6\times 6$ CHMs is also a long-standing open problem. Paper \cite{Butson} characterized the CHMs whose all elements are roots of unity of finite order, such as the Fourier matrices. Paper \cite{Taom} given the Tao  matrix which only consists of $1,\mathrm{e}^{\frac{2\pi \mathrm{i}}{3}}, \mathrm{e}^{\frac{4\pi \mathrm{i}}{3}}$, the Tao matrix does not belong to any parameterized matrices family. Paper \cite{BjrckGran} presented a trick whereby the faster algorithms for the homogeneous case can be used in the inhomogeneous case and found a 
$6\times 6$ CHMs family which does not belong to the Butson matrices by this algorithms.  
In 2011, Karlsson presented a three-parameter CHMs family in $\mathbb{C}^6$, he named this family “ the $H_2$-reducible matrices ”. As far as we know, most of known $6\times 6$ CHMs belong to the $H_2$-reducible matrices such as the Haagerup matrix \cite{Haagerup1997}, but the Tao matrix is an exception. Paper \cite{Sz12} presented 
a four-parameter $6\times 6$ CHMs family, but the analytic form of this  matrices is still unknown. More papers investigating the classification problem has been presented in \cite{2006Orthogonal,Tadej2006,Banica2009,Launey2001,DitaP2002,Hiranandani2014}. \par
In this paper, we investigate the eigenvalues and eigenvectors of CHMs. Specifically, we show that any $n\times n$ CHM with dephased form has two constant eigenvalues $\pm\sqrt{n}$ and has two eigenvectors $v_1=[1+\sqrt{n},1,...,1]^{\mathrm{T}}\in \mathbb{R}^n$ and $v_2=[1-\sqrt{n},1,...,1]^{\mathrm{T}}\in \mathbb{R}^n$ in Lemma \ref{firstl}. Further, If $\lambda \neq \pm \sqrt{n}$ is an eigenvalue of $H$ and $Hv=\lambda v$, where $v=\begin{bmatrix}
x_1\\
x_2\\
\vdots \\
x_n
\end{bmatrix}$, then $x_1=0$. We also show that if $\pm \sqrt{6},\lambda_1,...,\lambda_4$ are eigenvalues of a $6\times 6$ CHM $H$, then at most two of $\lambda_1,...,\lambda_4$ are the same in Theorem \ref{re2}. Finally, We show that there is no 6 × 6 CHM with four identical eigenvalues in Theorem \ref{re3}. A related case is the Hermitian CHMs with dephased form, its eigenvalues are $\sqrt{6},\sqrt{6},\sqrt{6},-\sqrt{6},-\sqrt{6},-\sqrt{6}$.
Our results might lead to the complete classification of $6\times 6$ CHMs.\par
The rest of this paper is structured as follows.  In Sec. \ref{sec:pre} we introduce some facts that used in this paper, such as the precise definition of MUBs and CHMs, some famous CHMs, some properties of CHMs and so on. In Sec. \ref{sec:eigofCHM}, we introduce the two main result of this paper. The first one is that if $\pm \sqrt{6},\lambda_1,...,\lambda_4$ are eigenvalues of a $6\times 6$ CHM $H$, then at most two of $\lambda_1,...,\lambda_4$ are the same. The second one is that there is no 6 × 6 CHM with four identical eigenvalues. We conclude in Sec. \ref{sec:con}.

 \section{Preliminaries}
 \label{sec:pre}
 In this section, we will introduce some facts that used in this paper. First of all, we review the precise definition of MUBs and CHMs.

 \subsection{MUBs and CHMs}
\label{subsec:MUBCHM}
The following two lemma are the precise definition of MUBs and CHMs. 
\begin{lemma}
    Suppose $\mathcal{A}_1={\{\ket{\psi_{1i}}\}}_{i=1,...,d}$,...,$\mathcal{A}_N={\{\ket{\psi_{Ni}}\}}_{j=1,...,d}$ are $N$ orthonormal basis in $\mathbb{C}^d$, if
\begin{eqnarray}
    |\braket{\psi_{jm}|\psi_{kn}}|=\frac{1}{\sqrt{d}},
\end{eqnarray}
for any $j\neq k,j,k=1,...,N$, $m,n=1,...,d$, then $\mathcal{A}_1$,...,$\mathcal{A}_N$ constitute $N$ MUBs.
\end{lemma}
\begin{lemma}
 If $H$ is an $n\times n$ complex matrix, and satisfies
\begin{enumerate}
\item all elements of $H$ have modulus one,
\item $HH^{\dagger}=n\rm{I}$,
\end{enumerate}
then $H$ is an $n\times n$ CHM.
\end{lemma}

If the set of $N$ MUBs in $\mathbb{C}^d$ exists and contains the identity, then one can show that any other matrix $U$ in the set satisfies that $\sqrt{n}U$ is a $n\times n$ CHM. Thus the CHMs is usually related to the MUBs.

Next, we introduce the complex equivalence of matrices.

\subsection{Complex equivalence of matrices}
\label{subsec:ce}

To find out the relation between different CHMs, we introduce the definition of the complex equivalence. We refer to the \textit{monomial unitary matrix} as a unitary matrix each of whose rows and columns has exactly one nonzero element. Then two matrices $U$ and $V$ are complex equivalent when $U=PVQ$ where $P,Q$ are monomial unitary matrices. In particular, when  $P,Q$ are both permutation matrices, $U$ and $V$ are equivalent. When  $P,Q$ are both real monomial unitary matrices, we say that $U$ and $V$ are real equivalent. One can verify that if $U$ and $V$ are equivalent or real equivalent then $U$ and $V$ are complex equivalent. 

Next, we list some famous CHMs and their eigenvalues.

\subsection{Some famous CHMs}
\label{subsec:chm}

The Tao matrix \cite{Taom}  has the following form
\begin{eqnarray}\label{eqn:2}
S_6^{(0)}=\begin{bmatrix}
1 & 1 & 1 & 1 & 1 & 1\\
1 & 1 & \omega & \omega & \omega^2 & \omega^2\\
1 & \omega & 1 & \omega^2 & \omega^2 & \omega \\
1 & \omega & \omega^2 & 1 & \omega & \omega^2 \\
1 & \omega^2 & \omega^2 & \omega & 1 & \omega \\
1 & \omega^2 & \omega & \omega^2 & \omega & 1
\end{bmatrix},
\end{eqnarray}
where $\omega=e^{\frac{2\pi i}{3}}$ or $e^{\frac{4\pi i}{3}}$. It is symmetric, isolated and belongs to the Butson matrices. The eigenvalues of the Tao matrix are $\sqrt{6}$ ,$\sqrt{6}$, $\dfrac{3+\sqrt{15}\mathrm{i}}{2}$, $\dfrac{3+\sqrt{15}\mathrm{i}}{2}$, $\dfrac{3-\sqrt{15}\mathrm{i}}{2}$, $\dfrac{3-\sqrt{15}\mathrm{i}}{2}$.

The Haagerup matrices \cite{Haagerup1997} are
\begin{eqnarray}\label{eq:H6q}
H_6^q=\begin{bmatrix}
1 & 1 & 1 & 1 & 1 & 1\\
1 & -1 & \mathrm{i} & \mathrm{i} & -\mathrm{i} & -\mathrm{i}\\
1 & \mathrm{i} & -1 & -\mathrm{i} & q & -q \\
1 & \mathrm{i} & -\mathrm{i} & -1 & -q & q \\
1 & -\mathrm{i} & \frac{1}{q} & -\frac{1}{q} & \mathrm{i} & -1 \\
1 & -\mathrm{i} & -\frac{1}{q} & \frac{1}{q} & -1 & \mathrm{i}
\end{bmatrix},
\end{eqnarray}
where $|q|=1$. The eigenvalues of the Haagerup matrices are $\sqrt{6},-\sqrt{6},\mathrm{i}-\sqrt{5},\mathrm{i}+\sqrt{5},-1-\sqrt{5}\mathrm{i},-1+\sqrt{5}\mathrm{i}$.

Suppose the $6\times6$ CHM $H_6$ is an {Hermitian} matrix, then $H_6=P H(t) P^\dag$, where $P$ is a complex permutation matrix, and $H(t)$ is the following Hermitian matrix \cite{2006Orthogonal}
\begin{eqnarray}
\label{eq:em}
H(t)=
\begin{bmatrix}
1& 1 & 1 & 1 & 1 & 1 \\
1&-1 & \frac{1}{x} & -y & -\frac{1}{x} & y \\
1& x & -1 & t & -t & -x \\
1& -\frac{1}{y}& \frac{1}{t} & -1 & \frac{1}{y} & -\frac{1}{t} \\
1& -x & -\frac{1}{t} & y & 1 & \frac{1}{z}\\
1& \frac{1}{y} & -\frac{1}{x} & -t & z & 1 \\
\end{bmatrix},
\end{eqnarray}
where $\theta \in [-\pi ,-arccos (\frac{-1+\sqrt{3}}{2})]\cup [arccos (\frac{-1+\sqrt{3}}{2}),\pi ]$, and the real parameters $x,y,z,t$ are given by
\begin{eqnarray}
&y=e^{i \theta}, z=\dfrac{1+2y-y^2}{y(-1+2y+y^2)}\\
&x=\dfrac{1+2y+y^2-\sqrt{2}\sqrt{1+2y+2y^3+y^4}}{1+2y-y^2}\\
&t=\dfrac{1+2y+y^2-\sqrt{2}\sqrt{1+2y+2y^3+y^4}}{-1+2y+y^2}.
\end{eqnarray}

The eigenvalues of the Hermitian matrix $H(t)$ are $\sqrt{6},\sqrt{6},\sqrt{6},-\sqrt{6},-\sqrt{6},-\sqrt{6}$. The Hermitian matrix with dephased form is the only $6\times 6$ CHM with dephased form whose eigenvalues only consist of two numbers.

 \section{Eigenvalues and eigenvectors of a CHM}
 \label{sec:eigofCHM}
 In this section, we will introduce our results in this paper. First, we give a lemma about the eigenvalues of CHMs.
 \begin{lemma}
 \label{firstl}
(i) If $H$ is an $n\times n$ unitary matrix, and $\lambda_1,...,\lambda_n$ are eigenvalues of $H$, then $|\lambda_k |=1$ for $k=1,...,n$.

(ii) If $H$ is a $6\times 6$ symmetric CHM, then there are real orthonormal vectors $\ket{v_1},...,\ket{v_6} $ such that $\ket{v_1},...,\ket{v_6}$ are eigenvectors of $H$.\qed
\end{lemma}

Next, we present the first result of this paper.
\begin{lemma}
\label {re1}
If $H$ is an $n\times n$ CHM with dephased form, then

(i) $H$ has two constant eigenvalues $\sqrt{n}$ and $-\sqrt{n}$, and has two eigenvectors $v_1=[1+\sqrt{n},1,...,1]^{\mathrm{T}}\in \mathbb{R}^n$ and $v_2=[1-\sqrt{n},1,...,1]^{\mathrm{T}}\in \mathbb{R}^n$. They satisfy
\begin{eqnarray}
Hv_1=\sqrt{n}v_1,\ Hv_2=-\sqrt{n}v_2.
\end{eqnarray}
where $\dfrac{1}{\sqrt{2n+2\sqrt{n}}}v_1, \dfrac{1}{\sqrt{2n-2\sqrt{n}}}v_2$ are orthonormal vectors.\\
(ii) If $\lambda \neq \pm \sqrt{n}$ is an eigenvalue of $H$ and $Hv=\lambda v$, where $v=\begin{bmatrix}
x_1\\
x_2\\
\vdots \\
x_n
\end{bmatrix}$, then $x_1=0$.\\
(iii) If $n\geq 4$ and $\sqrt{n},-\sqrt{n},\lambda_3,...,\lambda_n$ are eigenvalues of $H$, then $\lambda_3=\lambda_4=\cdots=\lambda_n$ is impossible.
\qed
\end{lemma}
\begin{proof}
(i)
Suppose
\begin{eqnarray}
H=\begin{bmatrix}
1 & 1 & 1 & \cdots & 1\\
1 & a_{22} & a_{23} & \cdots & a_{2n}\\
1 & a_{32} & a_{33} & \cdots & a_{3n}\\
\vdots & \vdots & \vdots & \ddots & \vdots\\
1 & a_{n2} & a_{n3} & \cdots & a_{nn}
\end{bmatrix}.
\end{eqnarray}
By $H^{\dagger}H=n\mathrm{I}$ we have $1+\sum\limits_{k=2}^na_{lk}=0\ (l=2,...,n)$. Therefore 
\begin{eqnarray}
H\begin{bmatrix}
1+\sqrt{n}\\
1\\
1\\
\vdots \\
1
\end{bmatrix}=\begin{bmatrix}
1+\sqrt{n}+1+1+\cdots+1\\
1+\sqrt{n}+\sum_{k=2}^na_{2k}\\
1+\sqrt{n}+\sum_{k=2}^na_{3k}\\
\vdots \\
1+\sqrt{n}+\sum_{k=2}^na_{nk}
\end{bmatrix}=\sqrt{n}\begin{bmatrix}
1+\sqrt{n}\\
1\\
1\\
\vdots \\
1
\end{bmatrix}.\nonumber \\
\end{eqnarray}
Hence $\sqrt{n}$ is an eigenvalue of $H$, $v_1$ is an eigenvector of $H$, and $Hv_1=\sqrt{n}v_1$. Similarly one can show that $-\sqrt{n}$ is an eigenvalue of $H$, $v_2$ is an eigenvector of $H$, and $Hv_2=-\sqrt{n}v_2$.\\
(ii) Since $v_1=\begin{bmatrix}
1+\sqrt{n}\\
1\\
1\\
\vdots \\
1
\end{bmatrix}$ and $v_2=\begin{bmatrix}
1-\sqrt{n}\\
1\\
1\\
\vdots \\
1
\end{bmatrix}$ are two constant eigenvectors of $H$, and $\lambda \neq \pm \sqrt{n}$ is an eigenvalue of $H$ and $Hv=\lambda v$. Therefore $\langle v | v_1 \rangle=\langle v | v_2 \rangle=0$, then we have $\langle v | v_1-v_2\rangle =\begin{bmatrix}
2\sqrt{n}x_1\\
0\\
0\\
\vdots \\
0
\end{bmatrix}$=0. Hence $x_1=0$.\\
(iii) Suppose $u_1=\dfrac{1}{\sqrt{2n+2\sqrt{n}}}\begin{bmatrix}
1+\sqrt{n}\\
1\\
1\\
\vdots \\
1
\end{bmatrix},u_2=\dfrac{1}{\sqrt{2n-2\sqrt{n}}}\begin{bmatrix}
1-\sqrt{n}\\
1\\
1\\
\vdots \\
1
\end{bmatrix},u_3,u_4,...,u_n$ are orthonormal eigenvectors of $H$ and $\lambda_3=\lambda_4=\cdots=\lambda_n$, then
\begin{eqnarray}
H&&=\sqrt{n}\ket{u_1}\bra{u_1}-\sqrt{n}\ket{v_2}\bra{v_2}+\lambda \sum_{k=3}^n\ket{v_k}\bra{v_k} \\
&&=\sqrt{n}\ket{v_1}\bra{v_1}-\sqrt{n}\ket{v_2}\bra{v_2}+\lambda (\mathrm{I}-\ket{v_1}\bra{v_1}-\ket{v_2}\bra{v_2})
\nonumber \\
\\
&&\small{=\dfrac{1}{n-1}\begin{bmatrix}
n-1 & n-1 & n-1 & n-1 & \cdots & n-1\\
n-1 & -1+(n-2)\lambda & -1-\lambda & -1-\lambda & \cdots & -1-\lambda \\
n-1 & -1-\lambda & -1+(n-2)\lambda & -1-\lambda & \cdots & -1-\lambda \\
\vdots & \vdots  & \vdots  & \vdots  & \vdots  & \vdots  \\
n-1 & -1-\lambda & -1-\lambda & -1-\lambda & \cdots & -1+(n-2)\lambda
\end{bmatrix}}.\nonumber \\
\end{eqnarray}

By the orthogonality of the second and third row vectors of $H$ one can show $H$ is not an $n\times n$ CHM. So $\lambda_3=\lambda_4=\cdots=\lambda_n$ is impossible, and we complete this proof.
\end{proof}

Now we will present our first main result of this paper.
\begin{theorem}
\label{re2}
Suppose $H$ is a $6\times 6$ CHM with dephased form. 

(i) $H$ is {an Hermitian matrix} if and only if $H$ has three equal eigenvalues and $\mathrm{Tr} H=0$, if and only if the eigenvalues of $H$ are $\sqrt{6}$ or $-\sqrt{6}$.

(ii) If $\sqrt{6},-\sqrt{6},\lambda_3,...,\lambda_6$ are eigenvalues of $H$, then at most two of $\lambda_3,...,\lambda_6$ are the same.

\end{theorem} 

\begin{proof}
(i) If $H$ is a $6\times 6$ Hermitian CHM with dephased form, one can verify that all eigenvalues of $H$ are $ \sqrt{6},\sqrt{6},\sqrt{6},-\sqrt{6},-\sqrt{6},-\sqrt{6}$ by equation \eqref{eq:em}. Hence there are three equal eigenvalues of $H$ and $\mathrm{Tr}(H)=0$.\par
If there are three equal eigenvalues of $H$ and $\mathrm{Tr}(H)=0$, then by Lemma \ref{lemma32} we have the eigenvalues of $H$ are $ \sqrt{6},\sqrt{6},\sqrt{6},-\sqrt{6},-\sqrt{6},-\sqrt{6}$, then by the spectral decomposition of $H$ we have 
\begin{eqnarray}
H=\sqrt{6}\sum_{k=1}^{3}\ket{u_k}\bra{u_k}-\sqrt{6}\sum_{k=4}^{6}\ket{u_k}\bra{u_k}.
\end{eqnarray}
Hence $H^{\dagger}=\sqrt{6}\sum_{k=1}^{3}\ket{u_k}\bra{u_k}-\sqrt{6}\sum_{k=4}^{6}\ket{u_k}\bra{u_k}=H$. So $H$ is an Hermitian matrix, we complete this proof.

(ii) Let $u_1=\dfrac{1}{\sqrt{12+2\sqrt{6}}}\begin{bmatrix}
1+\sqrt{6}\\
1\\
1\\
1\\
1\\
1
\end{bmatrix},u_2=\dfrac{1}{\sqrt{12-2\sqrt{6}}}\begin{bmatrix}
1-\sqrt{6}\\
1\\
1\\
1\\
1\\
1
\end{bmatrix},u_6=\begin{bmatrix}
0\\
a_0\\
a_1e^{\mathrm{i}t_1}\\
a_2e^{\mathrm{i}t_2}\\
a_3e^{\mathrm{i}t_3}\\
a_4e^{\mathrm{i}t_4}
\end{bmatrix}$ where $a_k\geq 0 (k=1,...,4)$. Suppose $u_1,u_2,...,u_6$ are orthonormal eigenvectors of $H$ and $\lambda_3=\lambda_4=\lambda_5=\lambda \neq \lambda_6$, then

\begin{eqnarray}
&&H=\sqrt{6}\ket{u_1}\bra{u_1}-\sqrt{6}\ket{u_2}\bra{u_2}+\lambda \sum_{k=3}^5\ket{u_k}\bra{u_k}+\lambda_6\ket{u_6}\bra{u_6}   \nonumber  \\ \nonumber
&&=\sqrt{6}(\ket{u_1}\bra{u_1}-\ket{u_2}\bra{u_2})+\lambda (\mathrm{I}-\ket{u_1}\bra{u_1}-\ket{u_2}\bra{u_2})+(\lambda_6-\lambda)\ket{u_6}\bra{u_6}\\ 
&&\small{=\dfrac{1}{5}\begin{bmatrix}
5 & 5 & 5 & 5 & 5 & 5\\
5 & -1+4\lambda & -1-\lambda & -1-\lambda & -1-\lambda & -1-\lambda \\
5 & -1-\lambda & -1+4\lambda & -1-\lambda & -1-\lambda & -1-\lambda \\
5 & -1-\lambda & -1-\lambda & -1+4\lambda & -1-\lambda & -1-\lambda \\
5 & -1-\lambda & -1-\lambda & -1-\lambda & -1+4\lambda & -1-\lambda \\
5 & -1-\lambda & -1-\lambda & -1-\lambda & -1-\lambda & -1+4\lambda
\end{bmatrix}+(\lambda_6-\lambda)\ket{u_6}\bra{u_6}.} \nonumber 
\end{eqnarray}
Let $f(x)=\frac{1}{5}(1-4\lambda )+x^2(\lambda_6-\lambda)$, $h(m,n)=\frac{1}{5}(-1-\lambda )+a_ma_ne^{\mathrm{i}(t_m-t_n)}(\lambda_6-\lambda)$. Suppose $t_0=0$, then $H=$
\begin{eqnarray}
\begin{bmatrix}
1 & 1 & 1 & 1 & 1 & 1\\
1 & f(a_0) & h(0,1) & h(0,2) & h(0,3) & h(0,4) \\
1 & h(1,0) & f(a_1) & h(1,2) & h(1,3) & h(1,4) \\
1 & h(2,0) & h(2,1) & f(a_2) & h(2,3) & h(2,4) \\
1 & h(3,0) & h(3,1) & h(3,2) & f(a_3) & h(3,4) \\
1 & h(4,0) & h(4,1) & h(4,2) & h(4,3) & f(a_4)
\end{bmatrix}.
\end{eqnarray}
Since elements of $H$ have modulus one, we have $|f(a_k)|=1(k=0,...,4)$, and $a_0,...,a_4$ are five roots of this equation. One can show that there are three equal numbers from $\{a_0,...,a_4 \}$. Without loss of generality, let $a_1=a_2=a_3$. Because elements of $H$ have modulus one and $\lambda \neq \lambda_6$, we have 
\begin{eqnarray}
|h(1,0)|=|h(2,0)|=|h(3,0)|,
\end{eqnarray}
namely
\begin{eqnarray}
|\frac{-1-\lambda}{5a_0a_1(\lambda_6-\lambda)}+e^{\mathrm{i}t_1}|=|\frac{-1-\lambda}{5a_0a_1(\lambda_6-\lambda)}+e^{\mathrm{i}t_2}|=|\frac{-1-\lambda}{5a_0a_1(\lambda_6-\lambda)}+e^{\mathrm{i}t_3}|.\nonumber\\
\end{eqnarray}
Suppose $\dfrac{-1-\lambda}{5a_0a_1(\lambda_6-\lambda)}=z$ and $z=|z|e^{-\mathrm{i}\theta}$, then we have
\begin{eqnarray}
||z|+e^{\mathrm{i}(t_1+\theta)}|=||z|+e^{\mathrm{i}(t_2+\theta)}|=||z|+e^{\mathrm{i}(t_3+\theta)}|.
\end{eqnarray}
Namely
\begin{eqnarray}
&&|z|^2+2|z|\cos (t_1+\theta)+1\\
=&&|z|^2+2|z|\cos (t_2+\theta)+1\\
=&&|z|^2+2|z|\cos (t_3+\theta)+1.
\end{eqnarray}
Hence
\begin{eqnarray}
\cos (t_1+\theta)=\cos (t_2+\theta)=\cos (t_3+\theta).
\end{eqnarray}
One can show that there are two equal numbers from $\{t_1,t_2,t_3 \}$. Without loss of generality, let $t_1=t_2$. Thus $h(1,0)=h(2,0)$, $h(1,3)=h(2,3)$, $h(1,4)=h(2,4)$. One can show that $H$ contains a $2\times 4$ submatrix with rank one under complex equivalence, hence $H$ is not a $6\times 6$ CHM. So $\lambda_3=\lambda_4=\lambda_5$ is impossible, and we complete this proof.
\end{proof}

Then, we will introduce the second main reslut of this paper.
\begin{theorem}
\label{re3}
There is no $6\times 6$ CHM with four identical eigenvalues.
\end{theorem}
\begin{proof}
    Suppose $H=[h_{jk}],(j,k=1,...,6)$ is a $6\times 6$ CHM. Up to a phase, we can assume that $\sqrt{6},\sqrt{6},\sqrt{6},\sqrt{6},\sqrt{6}\mathrm{e}^{\mathrm{i}a},\sqrt{6}\mathrm{e}^{\mathrm{i}b}$ are eigenvalues of $H$ where $a,b\in (0,\pi]$. By the spectral decomposition of $H$ we have
    \begin{eqnarray}
        H&&=\sqrt{6}(\proj{v_1}+\proj{v_2}+\proj{v_3}+\proj{v_4})+\sqrt{6}\mathrm{e}^{\mathrm{i}a}\proj{v_5}+\sqrt{6}\mathrm{e}^{\mathrm{i}b}\proj{v_6}\\
        &&\label{SPEH}=\sqrt{6}(\mathrm{I}- \proj{v_5}- \proj{v_6})+\sqrt{6}\mathrm{e}^{\mathrm{i}a}\proj{v_5}+\sqrt{6}\mathrm{e}^{\mathrm{i}b}\proj{v_6},
    \end{eqnarray}
    where $\ket{v_1},...,\ket{v_4}$ and $\ket{v_5}=\begin{bmatrix}
        g_0\\
        g_1\mathrm{e}^{\mathrm{i}s_1}\\
        g_2\mathrm{e}^{\mathrm{i}s_2}\\
        g_3\mathrm{e}^{\mathrm{i}s_3}\\
        g_4\mathrm{e}^{\mathrm{i}s_4}\\
        g_5\mathrm{e}^{\mathrm{i}s_5}
    \end{bmatrix}$, $\ket{v_6}=\begin{bmatrix}
        h_0\\
        h_1\mathrm{e}^{\mathrm{i}t_1}\\
        h_2\mathrm{e}^{\mathrm{i}t_2}\\
        h_3\mathrm{e}^{\mathrm{i}t_3}\\
        h_4\mathrm{e}^{\mathrm{i}t_4}\\
        h_5\mathrm{e}^{\mathrm{i}t_5}
    \end{bmatrix}$ are eigenvectors of $H$ and $g_k,h_k\in [0,1](k=0,...5),s_k,t_k\in (0,2\pi](k=1,...,5)$. Hence we have
    \begin{eqnarray}
        &&\label{sumgk}\sum\limits_{k=0}^5g_k^2=1,\\
        &&\label{sumhk}\sum\limits_{k=0}^5h_k^2=1,\\
        &&\label{gkhkei}g_0h_0+\sum\limits_{k=1}^5g_kh_k\mathrm{e}^{\mathrm{i}(s_k-t_k)}=0.
    \end{eqnarray}
    Let $s_0=t_0=0$, using \eqref{SPEH} one can verify that
    \begin{eqnarray}
    \label{case123}
        ||h_{jk}|-|h_{kj}||=|96g_jg_kh_jh_k\sin\dfrac{a}{2}\sin\dfrac{a-b}{2}\sin\dfrac{b}{2}\sin(s_j-s_k-t_j+t_k)|=0,
    \end{eqnarray}
    for $j,k=0,...,5$. By equation \eqref{case123} we have the following cases.\par
    Case 1. $g_jg_kh_jh_k=0,j,k,=0,...,5$.\par
    WLOG, let $g_0=0$, by Lemma \eqref{SPEH} one can verify that
    \begin{eqnarray}
        &&|h_{1k}|=\sqrt{6}|-1+\mathrm{e}^{\mathrm{i}b}|h_0h_k=1,k=1,...,5\\
        &&\label{hkk}|h_{kk}|=\sqrt{6}|1+(-1+\mathrm{e}^{\mathrm{i}a})g_k^2+(-1+\mathrm{e}^{\mathrm{i}b})h_k^2|=1,k=1,...,5
    \end{eqnarray}
    Hence $h_1=\cdots =h_5$. By \eqref{hkk} one can show that $g_2=g_3=g_4$, hence $H$ contains a $3\times 3$ submatrix with rank one, and it is impossible. 
    \par
    Case 2. $a=b$. By Lemma \eqref{SPEH} one can verify that
    \begin{eqnarray}
        h_{kk}=\sqrt{6}\bigg(1+(\mathrm{e}^{\mathrm{i}a}-1)(g_k^2+h_k^2)\bigg),k=1,...,6
    \end{eqnarray}
    Hence we have 
    \begin{eqnarray}
    \label{hkk=1}
        1=|h_{kk}|=6 (1 + 2 (-1 + (g_k^2+h_k^2)) (g_k^2+h_k^2)) - 2 (-1 + (g_k^2+h_k^2)) (g_k^2+h_k^2) \cos[a]),k=1,...,6
    \end{eqnarray}
    Using \eqref{hkk=1} we have
    \begin{eqnarray}
    \label{gkhk=1}
        g_k^2+h_k^2=\frac{1}{2}-\frac{\sqrt{9 \cos ^2a-3 \cos a-6}}{6 (1-\cos a)},
    \end{eqnarray}
    or
    \begin{eqnarray}
    \label{gkhk=2}
        g_k^2+h_k^2=\frac{1}{2}+\frac{\sqrt{9 \cos ^2a-3 \cos a-6}}{6 (1-\cos a)},
    \end{eqnarray}
    where $k=0,...,5$. One can show that
    \begin{eqnarray}
        0\leq \frac{\sqrt{9 \cos ^2a-3 \cos a-6}}{6 (1-\cos a)}\leq \frac{\sqrt{6}}{12},
    \end{eqnarray}
    If $\exists \  k$ such that $g_k^2+h_k^2=\dfrac{1}{2}+\dfrac{\sqrt{9 \cos ^2a-3 \cos a-6}}{6 (1-\cos a)}$, then 
    \begin{eqnarray}
        \sum_{k=1}^{6}(g_k^2+h_k^2)\geq \frac{6}{2}-\frac{(5-1)\sqrt{6}}{12}>2.
    \end{eqnarray}
    Thus 
    \begin{eqnarray}
    \label{gkhk=-}
        g_k^2+h_k^2=\frac{1}{2}-\frac{\sqrt{9 \cos ^2a-3 \cos a-6}}{6 (1-\cos a)},k=1,...,6
    \end{eqnarray}
    Using \eqref{sumgk},\eqref{sumhk} and \eqref{gkhk=-} one can verify that
    \begin{eqnarray}
    \label{csa,gk}
        \cos a=-\frac{7}{8},\sin a=\frac{\sqrt{15}}{8},g_k^2+h_k^2=\frac{1}{3}.
    \end{eqnarray}
    Let $g_k=\dfrac{\sqrt{3}}{3}\cos x_k,h_k=\dfrac{\sqrt{3}}{3}\sin x_k$, $x_k\in (0,\dfrac{\pi}{2})$, using \eqref{csa,gk} and \eqref{SPEH} we have
    \begin{eqnarray}
    \label{scvec}
        \sin 2 {x_u} \sin 2 {x_v} \bigg( \cos ({s_u}-{t_u})\cos ({s_v}-{t_v})+\sin ({s_u}-{t_u})\sin ({s_v}-{t_v})\bigg)+\cos 2x_u \cos 2x_v=-\frac{1}{5},
    \end{eqnarray}
    where $u,v=1,...,6,u\neq v$. \par
    Let $\ket{a_k}=\begin{bmatrix}
        \cos 2x_k\\
        \sin 2x_k \cos(s_k-t_k)\\
        \sin 2x_k \sin(s_k-t_k)
        \end{bmatrix}$, $k=1,...,6$. Obviously $\ket{a_k}$ is a unit vector and using \eqref{gkhkei} and \eqref{scvec} we have
        \begin{eqnarray}
            \braket{a_u|a_v}=-\frac{1}{5},
        \end{eqnarray}
        where $u,v=1,...,6,u\neq v$.
         Next, we construct the matrix $A=\begin{bmatrix}
             \ket{a_0} & \ket{a_1} &\ket{a_2} &\ket{a_3} &\ket{a_4} &\ket{a_5}
         \end{bmatrix}$. Then the matrix the rank of $A^{\dagger}A$ is at most three. On the other hand, 
         \begin{eqnarray}
             A^{\dagger}A=\begin{bmatrix}
                 1 & -\dfrac{1}{5}& -\dfrac{1}{5}& -\dfrac{1}{5}& -\dfrac{1}{5}& -\dfrac{1}{5} \\
                 -\dfrac{1}{5} & 1& -\dfrac{1}{5}& -\dfrac{1}{5}& -\dfrac{1}{5}& -\dfrac{1}{5} \\
                 -\dfrac{1}{5} & -\dfrac{1}{5}& -1& -\dfrac{1}{5}& -\dfrac{1}{5}& -\dfrac{1}{5} \\
                 -\dfrac{1}{5} & -\dfrac{1}{5}& -\dfrac{1}{5}& 1& -\dfrac{1}{5}& -\dfrac{1}{5} \\
                 -\dfrac{1}{5} & -\dfrac{1}{5}& -\dfrac{1}{5}& -\dfrac{1}{5}& 1& -\dfrac{1}{5} \\
                 -\dfrac{1}{5} & -\dfrac{1}{5}& -\dfrac{1}{5}& -\dfrac{1}{5}& -\dfrac{1}{5}& 1
             \end{bmatrix}.
         \end{eqnarray}
         One can verify that $\mathrm{rank}(A^{\dagger}A)=5$.
         So we have a contradiction, and the CHM in this case does not exist.
    \par 
    Case 3. $\sin(s_j-s_k-t_j+t_k)=0,j,k=0,...,5$. Let $s_j-t_j=N_j\pi,N_j=-1,0,1$. Hence there exists $\ket{v^{'}_5}=\begin{bmatrix}
        d_0\\
        d_1\\
        d_2\\
        d_3\\
        d_4\\
        d_5
    \end{bmatrix}$, $\ket{v^{'}_6}=\begin{bmatrix}
        f_0\\
        f_1\\
        f_2\\
        f_3\\
        f_4\\
        f_5
    \end{bmatrix}$ such that $\ket{v_1},...,\ket{v_4},\ket{v^{'}_5},\ket{v^{'}_6}$ are eigenvectors of $H$ and $d_k\in (0,1),f_k\in (-1,0)\cup(0,1),(k=0,...5)$. One can show that 
    \begin{eqnarray}
        &&\label{sumgk}\sum\limits_{k=0}^5d_k^2=1,\\
        &&\label{sumhk}\sum\limits_{k=0}^5f_k^2=1,\\
        &&\sum\limits_{k=0}^5d_kf_k=0.
    \end{eqnarray}
    By \eqref{SPEH} and $|h_{jk}|=1$ we have
    \begin{eqnarray}
        d_j^2d_k^2(1-\cos a)+f_j^2f_k^2(1-\cos b)+d_jd_kf_jf_k(1-\cos a-\cos b+\cos (a-b))=\frac{1}{12},
    \end{eqnarray}
    where $k\neq j,k,j=0,...,5$. 
    Let \begin{eqnarray}
    \ket{\alpha_{j}}=\begin{bmatrix}
        1\\
        d_j^2\\
        f_j^2\\
        d_jf_j
    \end{bmatrix},\ket{\beta_k}=\begin{bmatrix}
        -\dfrac{1}{12}\\
        (1-\cos a)d_k^2\\
        (1-\cos b)f_k^2\\
        (1-\cos a-\cos b+\cos (a-b))d_kf_k
    \end{bmatrix}.
    \end{eqnarray}
    Then for $k\neq j,k,j=0,...,5$, we have $\braket{\alpha_{j}|\beta_k}=0$. Thus $\ket{\beta_k}$ is orthogonal to $\ket{\alpha_j}$, where $j\neq k$. Hence one can show that the rank of the matrix 
    \begin{eqnarray}
        D=\begin{bmatrix}
        1 & 1 & 1 & 1 & 1 & 1\\
        d_0^2 &d_1^2 &d_2^2 &d_3^2 &d_4^2 &d_5^2\\
        f_0^2 &f_1^2 &f_2^2 &f_3^2 &f_4^2 &f_5^2\\
        d_0f_0 &d_1f_1 &d_2f_2 &d_3f_3 &d_4f_4 &d_5f_5
    \end{bmatrix}
    \end{eqnarray} is at most three. Suppose $\mathrm{rank}{D}=3$, without loss of generality, we may assume that the column vectors $\ket{\alpha_0},\ket{\alpha_1},\ket{\alpha_2}$ are linearly independent. So any one of the column vectors $\ket{\alpha_3},\ket{\alpha_4},\ket{\alpha_5}$ is in the space which spanned by $\ket{\alpha_0},\ket{\alpha_1},\ket{\alpha_2}$.
    Since $\braket{\alpha_{j}|\beta_k}=0$, one can show that $\ket{\alpha_3}=\ket{\alpha_4}=\ket{\alpha_5}$.
    By the same reason, $\ket{\alpha_3}$ is the same as one of $\ket{\alpha_0},\ket{\alpha_1},\ket{\alpha_2}$.
  Up to the permutation, we may assume that $\ket{\alpha_3}=\ket{\alpha_2}$. Hence $d_2=d_3=d_4=d_5,f_2=f_3=f_4=f_5$. Using \eqref{SPEH} one can verify that the rank of $\begin{bmatrix}
      h_{31} & h_{32}\\
      h_{41} & h_{42}\\
      h_{51} & h_{52}\\
      h_{61} & h_{62}
  \end{bmatrix}$ is one. Thus $H$ is not a CHM, and it is a contradiction. Similarly one can show that $\mathrm{rank}D<3$ is impossible. So the CHM in this case does not exist.
  \par
  By Case 1,2,3, we complete this proof.
\end{proof}

\section{Conclusions}
\label{sec:con}
In this paper, we investigated the eigenvalues and eigenvectors of CHMs. We obtained some interesting results such as the Lemma \ref{re1}, the Theorem \ref{re2} and the Theorem \ref{re3}. They showed some properties of the eigenvalues and eigenvectors of CHMs. So far, the probelm of characterizing the $6\times 6$ CHMs reached a bottleneck. We need to find some innovative method to investigate this problem. Perhaps the eigenvalues and eigenvectors of $6\times 6$ CHMs will be the critical clue. At the end of this paper, we present a conjecture about how to characterize the $6\times 6$ CHMs. We guess that using four eigenvalues as the parameters can construct a $6\times 6$ CHMs family with dephased form, such that any two CHMs $A_1,A_2$ of this family satisfy $A_1=PA_2P^{\dagger}$ or $A_1=PA_2^{\mathrm{T}}P^{\dagger}$ where $P$ is a monomial unitary matrix, and the whole CHMs consist of the four parameters family.
\section*{Acknowledgements}

All authors were supported by the NNSF of China (Grant No.11871089), and the Fundamental Research Funds for the Central Universities (Grant No.ZG216S1902).

\section*{Data Availability Statements}
The data that support the findings of this study are available from the corresponding author upon reasonable request.

\section*{Competing Interests Statement}
The authors declare that there is no conflict of interest.

\bibliographystyle{unsrt}

\bibliography{mengfan}
\end{document}